\documentclass[
 aip, jap,
 amsmath,amssymb,
 reprint
]{revtex4-1}

\usepackage{graphicx}
\usepackage{dcolumn}
\usepackage{bm}
\usepackage{xcolor}
\usepackage[utf8]{inputenc}
\usepackage[T1]{fontenc}
\usepackage{mathptmx}
\usepackage{etoolbox}
\usepackage{orcidlink}

\usepackage{hyperref}
\hypersetup{colorlinks=true, linkcolor=blue, citecolor=blue, urlcolor=blue}

\makeatletter
\def\@emails#1#2#3#4{%
  \endgroup
  \patchcmd{\titleblock@produce}
    {\frontmatter@RRAPformat}
    {\frontmatter@RRAPformat
      {\produce@RRAP{*#1\href{mailto:#2}{#2}}}
     \frontmatter@RRAPformat
      {\produce@RRAP{*#3\href{mailto:#4}{#4}}}
    }{}{}
}

\newcommand{\iitrCPQCT}{Centre for Photonics and Quantum Communication Technology, Indian Institute of Technology Roorkee, Roorkee 247667, India}

\newcommand{\iiserps}{Department of Physical Sciences, Indian Institute of Science Education and Research Berhampur, Berhampur 760003, India}

\newcommand{\iitrPH}{Department of Physics, Indian Institute of Technology Roorkee, Roorkee 247667, India}

\begin{document}

\preprint{AIP/123-QED}

\title[Quantum Thermal Field Effect Transistor]{Quantum Thermal Field Effect Transistor}

\author{Abhijeet Kumar}
 \affiliation{\iiserps}
\author{Soniya Malik\,\orcidlink{0009-0009-6954-9767} }
\email{soniya@cpqct.iitr.ac.in}
\affiliation{\iitrCPQCT}

\author{P. Arumugam \,\orcidlink{0000-0001-9624-8024}}
\email{arumugam@ph.iitr.ac.in}
\affiliation{\iitrCPQCT}
\affiliation{\iitrPH}

\date{\today}

\begin{abstract}
We propose and analyse a quantum thermal field-effect transistor (qtFET) composed of left-qubit, middle-qutrit, and right-qubit subsystems. In this architecture, the left qubit is coupled to the middle qutrit, which in turn interacts with the right qubit. Each subsystem interacts independently with its respective baths. The middle subsystem serves as a modulator. We have shown that the qtFET exhibits functionality analogous to that of a conventional electronic field-effect transistor (eFET). The left, right, and middle subsystems of the qtFET correspond to the drain, source, and gate of an eFET in a common gate configuration, respectively. Our results show that the qtFET can precisely modulate thermal currents, highlighting its potential as a fundamental building block for quantum thermal devices and amplifiers in emerging quantum technologies. 
\end{abstract}

\maketitle

\section{\label{sec:level1}Introduction}
Electronic devices with dimensions approaching the atomic scale are being made. In these devices, heat dissipation is a major contributor to the failure of Moore's law~\cite{Schaller1997MooresLP, Krishnan2007ThermalMoore}. To overcome this limitation, quantum thermal devices are being investigated. Recently, there has been extensive research in quantum thermal devices and nanotechnology~\cite{Thermal2008Chung,Peres1995Quantum,Franklin2010LengthScaling,Franklin2022Carbon,Zhang2013Nanomaterials,Pomerantseva2019Energy,Curto2010Unidirectional,Gunathilake2025Advanced,FadakarMasouleh2016NanoStructured,Das2016NanoStructuredGaAs,Pekola2015Towards,Vinjanampathy2016Quantum,Sothmann2015Thermoelectric,Arrachea2023Energy,Blok2025Quantum,Wang_2008,PhysRevE.109.014142}. Quantum thermal devices can be used to regulate heat dissipation and to use waste thermal energy for useful work. Additionally, quantum thermal devices can be utilized in quantum computers to minimize the thermal noise and maintain the system's temperature stability. It is also possible to perform quantum computations using thermal logic gates~\cite{PhysRevLett.99.177208}. Nearly $63\%$ of global primary energy across all sectors is lost as thermal energy during combustion and heat transfer processes~\cite{FORMAN20161568}. Thermal management at the quantum scale~\cite{BenAbdallah2017Thermotronics} can resolve this global crisis, which can subsequently reduce global warming~\cite{Sun2023_TESroles,Ononogbo2023_WHR}. Quantum thermal transistors~\cite{Joulain_2016,gpmp-clgt,Guo_2018,Du2019Quantum,Liu2022CommonEnvironmental} based on bipolar junction transistors (BJTs) and quantum thermal diodes~\cite{rajapaksha2024enhanced} have been investigated. The quantum thermal diode model proposed by Rajapaksha \textit{et al.} showed enhanced thermal rectification by considering the system consisting of a qutrit and a qubit connected to heat baths, which operates without any additional sources to enable transitions between the qubit and the qutrit~\cite{rajapaksha2024enhanced}. Joulain \textit{et al.} proposed a quantum thermal transistor made up of three two-level subsystems modeled based on a BJT, where each subsystem is coupled to a corresponding thermal bath ~\cite{Joulain_2016}. Also, thermal rectifiers and thermal transistors have been investigated~\cite{PhysRevLett.93.184301,PhysRevLett.95.104302,PhysRevLett.97.124302,PhysRevB.76.020301, Hu_2009,Pereira_2011,Segal2005SpinBoson}. Thermal machines have also been experimentally investigated~\cite{Senior2020HeatRectification,Ronzani2018Tunable,Rossnagel2016SingleAtom}. A quantum thermal transistor enables controlled manipulation and amplification of heat currents at the quantum scale, analogous to how an electronic transistor regulates electrical currents. Quantum thermal transistors composed of qubit and qutrit subsystems have been investigated in the case of BJTs~\cite{PhysRevE.109.064146,Hioe1982Nonlinear,5x8m-bhgd}.

In this paper, a quantum thermal field-effect transistor (qtFET) is proposed; each consecutive subsystem interacts with the next in a chain-like manner, analogous to an electronic field-effect transistor (eFET), where the source is connected to the drain via a gate terminal between them. The qtFET is composed of three subsystems, with the right and the left subsystems corresponding to the source and drain, respectively. The middle subsystem is equivalent to the gate terminal. It is demonstrated that the qtFETs exhibit a relationship between the thermal current and the temperature difference, analogous to the relationship between the current and the voltage of an eFET. The complete description of the quantum thermal machine model we have proposed is described in Sec.~\ref{sec:level2}. In Sec.~\ref{sec:level3}, we discuss the results of our proposed model of a quantum thermal machine and show the analogy between our model and an eFET. The conclusion and future directions are presented in Sec.~\ref{sec:level4}.
\begin{figure}
    \centering
    \includegraphics[width=0.46\linewidth]{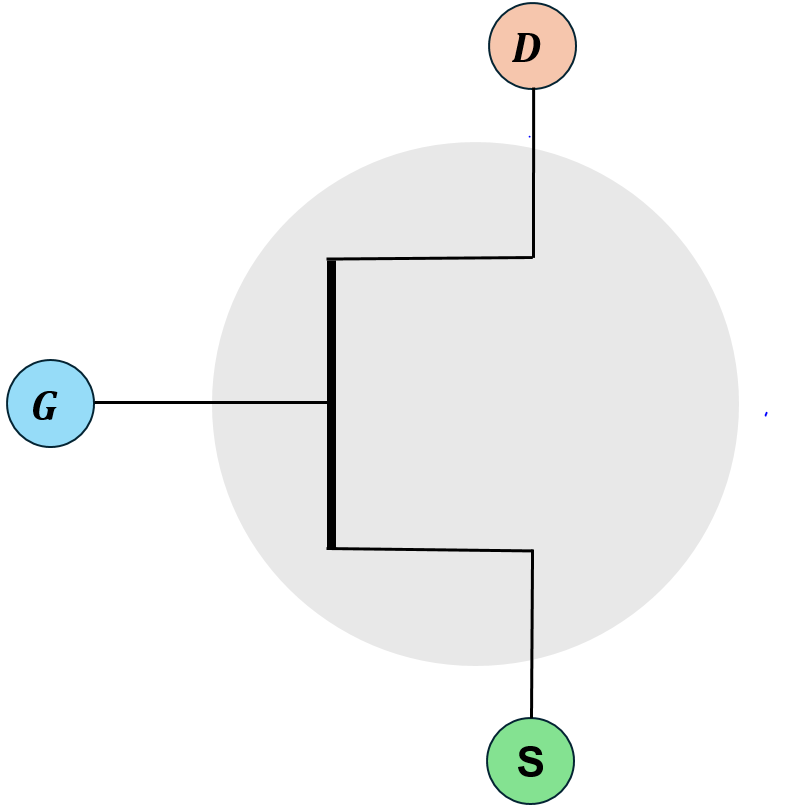}
    \includegraphics[width=0.48\linewidth]{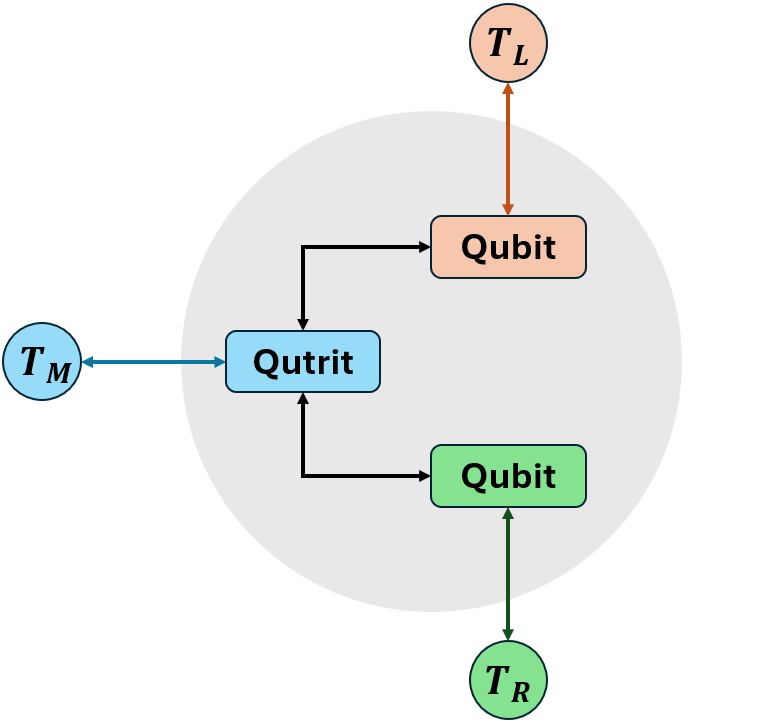}
    \caption{An electronic field-effect transistor eFET (left) and the proposed qtFET (right).}
    \label{fig:EFET}
\end{figure}

\section{\label{sec:level2}MODEL AND DYNAMICS}
An eFET is an electronic device used to control the flow of electric current in circuits by applying an electric field in a semiconductor channel (see Fig.~\ref{fig:EFET}). The eFETs are voltage-controlled devices where the current through them is regulated by a gate voltage.

The proposed qtFETs comprise a qubit-qutrit-qubit system with nearest-neighbor coupling, where each subsystem is coupled individually to a heat bath (see Fig.~\ref{fig:EFET}). The subscripts $L$, $M$, and $R$ are used for the left, middle, and right subsystems, as well as for the heat baths.

 \begin{figure}[h]
    \centering
    \includegraphics[width=0.98\linewidth]{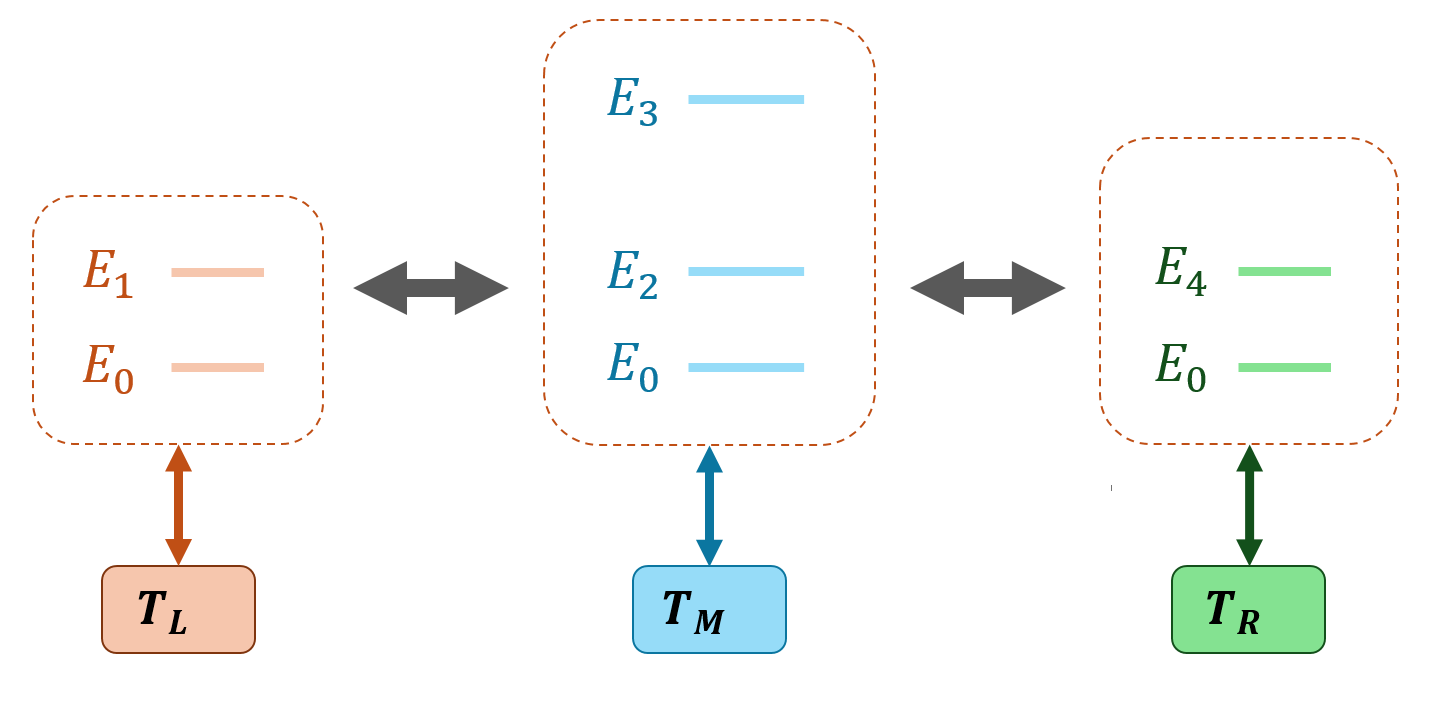}
    \caption{Quantum thermal field effect transistor energy levels.}
    \label{fig:Elevel}
\end{figure}

A qubit is a two-level quantum system having computational basis states $\vert0_{2}\rangle = [1,0]^{T}$ and $\vert1_{2}\rangle = [0,1]^{T}$. Similarly, a qutrit is a three-level quantum system with computational basis states $\vert0_{3}\rangle = [1,0,0]^{T},\vert1_{3}\rangle =[0,1,0]^{T} $ and $\vert2_{3}\rangle = [0,0,1]^{T}$. The subscripts $2$ and $3$ are used as shorthand notation for qubits and qutrits, respectively. The left qubit has energy eigenvalues $E_0$ and $E_1$, the middle qutrit has energy eigenvalues $E_0$, $E_2$ and $E_3$ and the right qubit has energy eigenvalues $E_0$ and $E_4$~\cite{Bohr1913,Griffiths2018_QMReview,Planck1901} (see Fig.~\ref{fig:Elevel}). Assuming the ground state energy $E_0$ to be $0$, the Hamiltonian for the system, according to Sakurai and Napolitano \cite{sakurai2017modern}, can be written as
\begin{equation}
    \begin{aligned}
        H_{2}^{L}  &= E_{1}\vert1_{2}\rangle\langle1_{2}\vert,\\
        H_{3}^{M}&= E_{2}\vert1_{3}\rangle\langle1_{3}\vert+E_{3}\vert2_{3}\rangle\langle2_{3}\vert, \\ 
        H_{2}^{R}& =E_{4}\vert1_{2}\rangle\langle1_{2}\vert.
    \end{aligned}
\end{equation}
Thus, the free Hamiltonian (without any interaction between the qubits and qutrits) of the system is
\begin{equation}
H_{0}=H_{2}^{L}\otimes I_{3}\otimes I_{2}+I_{2}\otimes H_{3}^{M}\otimes I_{2}+I_{2}\otimes I_{3}\otimes H_{2}^{R}.
\end{equation}
We define the transition operator for the two-level subsystem for the transition from $1^{\text{st}}$ to $0^{\text{th}}$ level as
\begin{equation}
    \sigma_{-}=\vert0_{2}\rangle\langle1_{2}\vert,
\end{equation}
the transition operator for the two-level subsystem for the transition from $0^{\text{th}}$ to $1^{\text{st}}$ level as
\begin{equation}
    \sigma_{+}=\vert1_{2}\rangle\langle0_{2}\vert,
\end{equation}
and the transition operators for the three-level subsystem as
\begin{align}
O_{01} =\vert 0_{3}\rangle\langle 1_{3}\vert, \quad
O_{12} =\vert1_{3}\rangle\langle2_{3}\vert, \nonumber\\
O_{01}^{\dagger}=\vert 1_{3}\rangle\langle 0_{3}\vert, \quad  O_{12}^{\dagger}=\vert 2_{3}\rangle\langle 1_{3}\vert.
\end{align}

The left subsystem interacts with the middle subsystem, which contributes a term to the total system Hamiltonian with strength $g_{LM}$. The interaction Hamiltonian between the left qubit and the middle qutrit is
\begin{align}
H_{I}^{LM}=g_{LM}\left(\sigma_+\otimes O_{01}\otimes I_{2}+\sigma_-\otimes O_{01}^\dagger \otimes I_{2}\right).
\end{align}
Similarly, the Hamiltonian for the interaction between the middle subsystem and the right subsystem with strength $g_{MR}$ is given by
\begin{align}
H_{I}^{MR}=g_{MR}\left( I_{2}\otimes O_{01}\otimes\sigma_++I_{2}\otimes O_{01}^\dagger \otimes\sigma_- \right).
\end{align}
The total Hamiltonian of the system is
\begin{align}
        H=H_{0} +H_{I}^{LM}+H_{I}^{MR}.
 \end{align}

Using Born-Markov approximation and secular approximation, the system's evolution is modelled based on the Lindblad master equation as
\begin{equation}
\frac{d\rho}{dt}=-i\left[H,\rho\right]+\mathcal{D}_{L}[\rho]+\mathcal{D}_{R}[\rho]+\mathcal{D}_{M1}[\rho]+\mathcal{D}_{M2}[\rho].
\end{equation} 

Here the thermal energy dissipates from each bath and the corresponding dissipators are
\begin{equation}
\mathcal{D}_P[\rho] = \kappa_P\left( n_B \mathcal{L}[P^\dagger] \rho  + (n_B+1) \mathcal{L}[P] \rho\right),
\end{equation}
with the dissipation rate $\kappa_P$  and the Lindblad superoperator
\begin{equation}
\mathcal{L}[P]\rho = P\rho P^\dagger -\frac{1}{2}\left\{ P^\dagger P , \rho\right\},    
\end{equation}
and 
\begin{equation}
    n_B = \frac{1}{\exp\{\Delta E/k_B T_P\}-1},
\end{equation}
is the occupation number of the bath $P$ at temperature $T_P$. $\Delta E$ is the transition energy associated with the operator $P$. Here, $\hbar$ and $k_{B}$ are taken in natural units. 
$\kappa_{L},\kappa_{M}$, and $\kappa_{R}$ represent the dissipation rates between the left bath and the qubit, the middle bath and the qutrit, and the right bath and the qubit, respectively.

The thermal current flows from the bath to the corresponding subsystems. $J_{L}$, $J_{M}$, and $J_{R}$ stand for the flow of thermal currents from the left bath to the left subsystem, the middle bath to the middle subsystem, and the right bath to the right subsystem, respectively.  The flow of thermal current from the left bath into the left subsystem, according to the definition used by Alicki~\cite{RAlicki1979} is
\begin{equation}
J_{L}=-\text{Tr}\left(H\mathcal{D}_{L}[\rho]\right).
\end{equation}
Similarly, the flow of thermal current from the right and the middle bath into the right and the middle subsystem, respectively, is
\begin{equation}J_{R}=-\text{Tr}\left(H\mathcal{D}_{R}[\rho]\right),\end{equation}
\begin{equation}J_{M}=-\text{Tr}\left(H\mathcal{D}_{M}[\rho]\right).\end{equation}

\section{\label{sec:level3}RESULTS}

The left, middle, and right subsystems of the qtFET can be mapped onto the drain, gate, and source, respectively, of an eFET. Analogously, the middle bath temperature ($T_M$) plays the role of a control parameter, similar to the gate voltage in an eFET. This correspondence establishes a thermal analogue of transistor action, where the heat current is modulated in a manner analogous to charge transport in electronic devices. 

\begin{figure}
    \centering
    \includegraphics[width=0.95\linewidth]{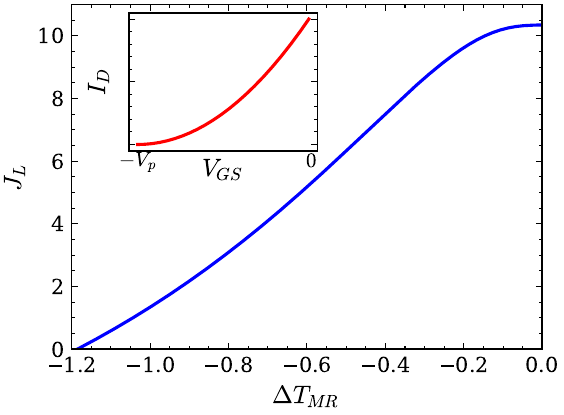}
    \caption{Thermal current $J_{L}$ vs.\ $\Delta T_{MR}$ of qtFET and $I_D$ vs. $\Delta V_{GS}$ for eFET (inset).
    Parameters are taken as $E_{1}=E_{2}=E_{4}=\omega_{0}$, $E_{3}=3\omega_{0}$, $g_{LM}=g_{MR}=0.1\omega_{0}$, 
    $\kappa_{L}=\kappa_{R}=0.05\omega_{0}$, $\kappa_{M}=0.02\omega_{0}$, $T_{L}=2.0\omega_{0}$, and $T_M=0.1\omega_{0}$, with $\omega_0$ as the energy spacing between the levels of the qubit.}
    \label{fig:JLvsTMR}
\end{figure}

Fig.~\ref{fig:JLvsTMR} shows the variation of thermal current $J_{L}$ as a function of the temperature difference between the middle and right baths ($\Delta T_{MR}$). The observed behaviour resembles the drain current ($I_{D}$) versus gate-to-source voltage ($V_{GS}$)  characteristic curve of an eFET. In both curves, the current exhibits a quadratic dependence on the control parameter. Moreover, the thermal current approaches zero below a certain threshold value of $\Delta T_{MR}$, which is directly analogous to the cut-off voltage in an eFET. This indicates the presence of a thermal threshold condition required to initiate current flow, reinforcing the gate-like control exerted by $T_{M}$.

\begin{figure}
    \centering
    \includegraphics[width=0.95\linewidth]{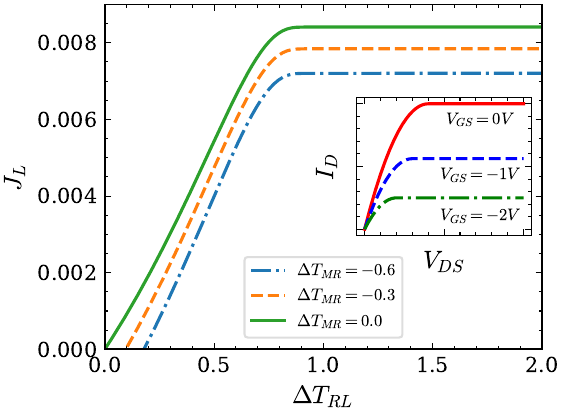}
    \caption{Thermal current $J_{L}$ vs. $\Delta T_{RL}$ different values of $\Delta T_{MR}$ of qtFET and $I_D$ vs. $V_{DS}$ different values of $V_{GS}$ of eFET (inset). Parameters are considered to be $E_{1}=E_{2}=E_{4}=\omega_{0},E_{3}=3\,\omega_{0},\;g_{LM}=g_{MR}=0.1\,\omega_{0},$ $\kappa_{L}=\kappa_{R}=0.05\,\omega_{0},\kappa_{M}=0.02\,\omega_{0}.$}
    \label{fig:JLvsTRL}
\end{figure}

Fig.~\ref{fig:JLvsTRL} represents the dependence of $J_{L}$ on the temperature difference between the left and right bath $(\Delta  T_{RL})$. This behaviour is analogous to $I_{D}$ versus drain-to-source voltage ($V_{DS}$) characteristics of an eFET. Initially, the thermal current increases with $\Delta T_{RL}$, reflecting a quadratic-like regime similar to the linear region of an eFET. A further increase in $\Delta T_{RL}$, $J_L$ tends towards a saturation value, mirroring the saturation region of an eFET, where the drain current has weak dependence on $V_{DS}$. This saturation behaviour indicates that the heat transport becomes limited by intrinsic system parameters rather than the applied thermal bias.

Moreover, the family of curves obtained by plotting $J_{L}$ versus $ \Delta T_{RL}$ for different values of $\Delta T_{MR}$ closely resembles the output characteristic, i.e., $I_{D}$ versus $V_{DS}$ of an eFET for different $V_{GS}$ values.
A decrease in $\Delta T_{MR}$ leads to the systematic reduction in $J_L$, analogous to the reduction of drain current with decreasing $V_{GS}$.This clearly demonstrates that $T_{M}$ effectively controls the thermal current. The ability to modulate current with an external control parameter is a key transistor property demonstrated here. The strong similarity between these characteristic curves confirms that the qtFET reproduces the essential operational features of an eFET in the thermal domain.

\begin{figure}[h!]
    \centering
    \includegraphics[width=1\linewidth]{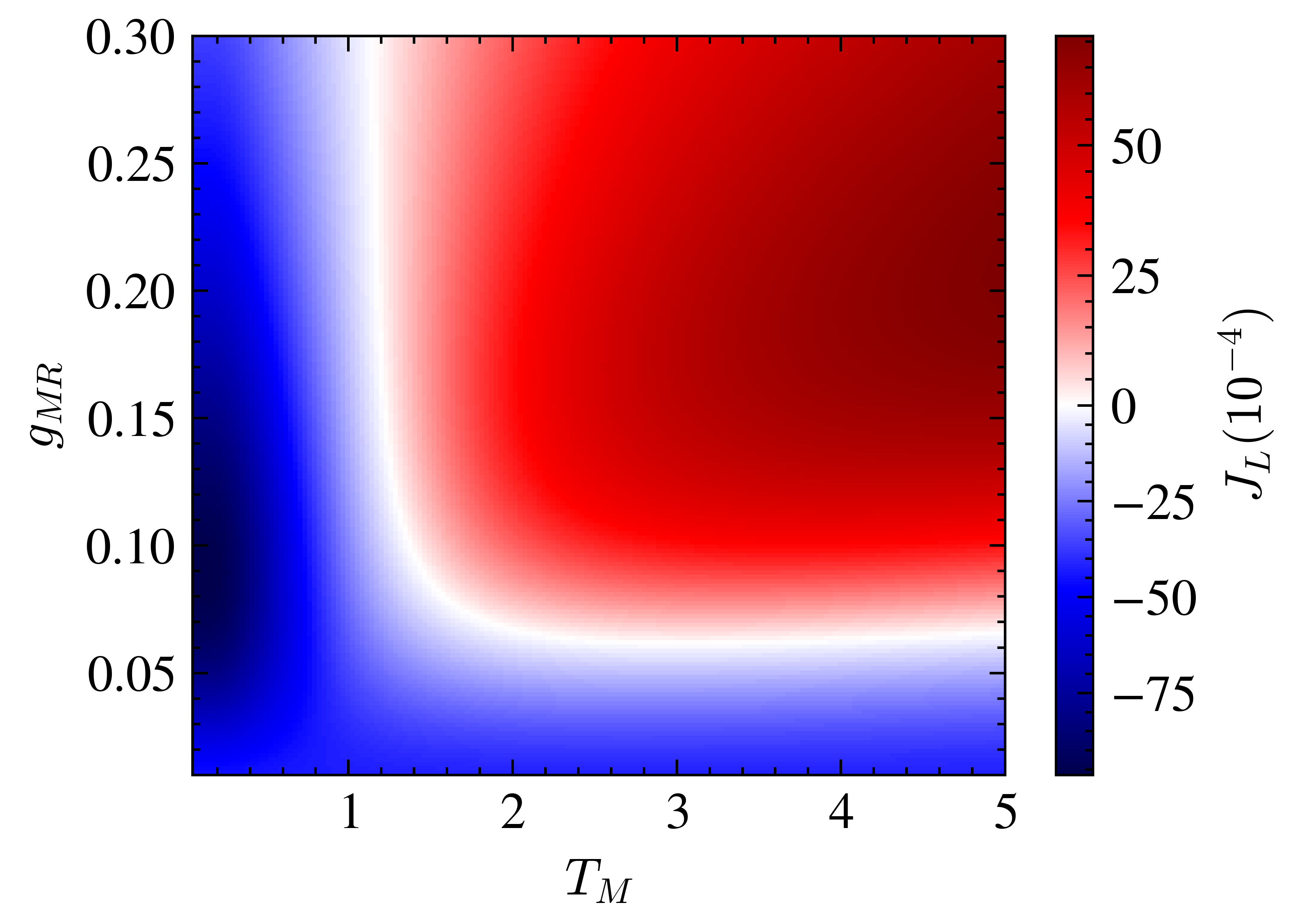}
    \caption{Variation of thermal current $J_{L}$ with respect to interaction strength $g_{MR}$ and right bath temperature $T_{M}$ for $E_{1}=E_{2}=E_{4}=\omega_{0},E_{3}=3\omega_{0},\;T_{L}=\omega_{0},T_{R}=0.1\omega_{0},g_{LM}=0.05\omega_{0},$$\kappa_{L}=\kappa_{R}=0.05\omega_{0},\kappa_{M}=0.002\omega_{0}.$}
    \label{fig:6}
\end{figure}

Fig.~\ref{fig:6} represents the variation of the left bath thermal current ($J_{L}$) as a function of the interaction strength between the middle qutrit and right qubit ($g_{MR}$) and the middle bath temperature ($T_{M}$). For the interaction strength range from $0\,\omega_{0}$ to $0.05\,\omega_{0}$, the left bath thermal current varies within a negative current range, which means the thermal current is flowing from the left subsystem into the left bath which is analogous to negative differential thermal resistance region of an eFET. Then, for the interaction range from $0.05\,\omega_{0}$ to $0.10\,\omega_{0}$, the thermal current starts varying from the negative thermal current and then reaches the minimum zero current, which is similar to the cutoff voltage in eFET. Also, when the interaction strength is less than $0.05\,\omega_{0}$, then there is only a slight variation in $J_{L}$ showing weak dependence of $J_L$ on $T_M$ for smaller values of $g_{MR}$. But $J_L$ depends significantly on the $T_M$  as the $g_{MR}$ becomes more than $0.10\,\omega_{0}$. The thermal current varies significantly even in response to slight variations in the middle bath temperature for the interaction strength range of $0.10\,\omega_{0}$ to $0.30\,\omega_{0}$. This confirms that $T_M$ acts as a modulator to regulate thermal current $J_{L}$ using optimised parameters, which is analogous to an eFET in which gate-to-source voltage acts as a modulator to regulate drain current.

\section{\label{sec:level4}CONCLUSION}
A quantum system comprising left qubit, middle qutrit, and right qubit subsystems can be realised as a Field-Effect Transistor (FET) when connected to optimised heat baths, yielding a quantum thermal FET (qtFET) that can be operated analogously to an electronic FET (eFET). As eFETs are integral components of low-noise amplifiers, they are ideal for weak-signal amplification~\cite{razavi2001design}. The qtFETs could help modulate thermal currents to maintain thermal stability. It can control thermal current using quantum mechanical states, and advanced thermal management is crucial for electronics~\cite{gpmp-clgt}. Quantum thermal transistors have numerous applications ~\cite{Cavaliere2023Hybrid} in quantum technology, with the potential to revolutionise the field of quantum computing~\cite{PhysRevE.99.032112}. The qtFETs should be further explored along similar lines to enhance their applications. Optimising the parameters of qtFETs with specialised algorithms and also automating the process of parameter optimisation using quantum or hybrid algorithms will make the qFETs experimentally realisable on various experimental platforms such as superconducting circuits~\cite{PhysRevB.101.184510}, ultracold atoms,~\cite{PRXQuantum.2.030310} and continuous variable electromechanical systems~\cite{article}.

\section{ACKNOWLEDGMENTS}
We acknowledge the support by the Science and Engineering Research Board (SERB) under Grant Code: CRG/2022/009359. Abhijeet Kumar acknowledges the Department of Science and Technology (DST), Government of India, for financial support through the INSPIRE Scholarship for Higher Education (SHE) under registration number 21002. 

\providecommand{\noopsort}[1]{}\providecommand{\singleletter}[1]{#1}%

\end{document}